\documentstyle[sprocl]{article}

\def\be{\begin{equation}}
\def\ee{\end{equation}}
\def\bea{\begin{eqnarray}}
\def\eea{\end{eqnarray}}
\renewcommand{\*}{ \hspace{-6pt}&=&\hspace{-6pt} }

\begin{document}
\hspace*{\fill} NSF-ITP-99-101\\
\hspace*{\fill} hep-th/9908202
\title{SUPERSYMMETRIC QUANTUM MECHANICS\\ FROM LIGHT CONE 
QUANTIZATION}

\author{ S. HELLERMAN }

\address{Department of Physics, UCSB, Santa
Barbara, CA 93106\\ \rm
sheller@twiki.physics.ucsb.edu}

\author{ J. POLCHINSKI }

\address{Institute for Theoretical Physics, UCSB, Santa Barbara,
CA 93106-4030\\ \rm  joep@itp.ucsb.edu}
\centerline{}

\maketitle\abstracts{We study the supersymmetric quantum
mechanical systems that arise from discrete light cone
quantization of theories with minimal supersymmetry in various
dimensions.  These systems which have previously arisen in the
study  of black hole moduli spaces, are distinguished by having
fewer fermionic fields than the familiar K\"ahler and hyper-K\"ahler
models. This is a contribution to the Yuri Golfand memorial
volume.}

\section{Preface}

In the three decades since the pioneering
works,\cite{susy} supersymmetry has steadily grown both in
importance and in depth.  It is the most widely
anticipated form of new physics to be seen near the weak scale;
it is central to the structure of string theory; it is the key
to the recent understanding of nonperturbative physics both in 
field theory and string theory; it is the focal point of many
connections between mathematics and physics; and, it appears to be
the principle that is responsible for cancellation of quantum
corrections, stabilizing both the weak symmetry breaking scale and
spacetime itself.  It seems likely that we still have much to
learn in these and other directions.

Even in quantum mechanics, supersymmetry has many applications.
Most recently, through discrete light cone quantization,
supersymmetric quantum mechanics has been proposed as a
nonperturbative formulation of M theory and other exotic systems.
Supersymmetric quantum mechanics is in some ways less
constrained than supersymmetric field theory, due to the
reduced spacetime symmetry.  For example, there is no direct
relation between the number of bosonic and fermionic fields
(quantum mechanical coordinates). In this paper we study some
supersymmetric quantum mechanical systems that arise from the
discrete light cone quantization of theories with minimal
supersymmetry in various dimensions.  These are distinguished by
having fewer fermionic fields than the familiar K\"ahler and
hyper-K\"ahler models.
 
\section{Introduction}

Light-cone quantization \cite{dirac} has played an important role
in field and string theory.  In this description, free quanta
satisfy nonrelativistic kinematics, with the positive `spatial'
momentum $p_-$ playing the role of the mass.  Of particular
interest is discrete light-cone quantization (DLCQ),\cite{dlcq}
in which the light-like spatial direction is compact.  In this
case $p_-$ is quantized as well as positive, so that a sector of
given total $p_-$ contains only a finite number of nonrelativistic
particles.  This has been used for numerical and analytic study of
quantum field theories, and recently has been used to provide a
nonperturbative definition of M theory.\cite{matrix}

In recent work \cite{hp1,hp2} we have attempted to employ DLCQ to
study weak/strong duality in supersymmetric gauge
theories.  This has required a version of DLCQ that preserves
this duality, the so-called light-like limit (LLL).
This is more complicated than the usual DLCQ, in that one must
in many cases solve a dynamical problem just to obtain the DLCQ
Hamiltonian: it is an effective Hamiltonian in the Wilsonian
sense, rather than a simple reduction of the original
Hamiltonian.

Given this complication, it is useful to take maximum advantage
of supersymmetry to restrict the form of the Hamiltonian.  We
are thus interested in various supersymmetric quantum mechanical
systems.  In this paper we focus on field theories with minimal
(unextended) supersymmetry in various dimensions, because these
lead to supersymmetric QM systems that are somewhat unfamiliar.
The
more general context of DLCQ and LLL will be discussed in a
forthcoming paper.\cite{hp2}

Let us begin with some simple
counting. We start with a field theory in $d$ spacetime
dimensions with ${\cal N}$ real supersymmetry charges.  The
nonrelativistic particles move in the $d-2$ transverse dimensions.  A
sector with
$k$ particles is then described by quantum mechanics with
$N_B = k(d-2)$ real bosonic coordinates.

The supersymmetries separate into 
$\frac{1}{2}{\cal N}$ dynamical charges $Q_\alpha$ and 
$\frac{1}{2}{\cal N}$ kinematical charges $q_\alpha$, with the
algebra
\begin{eqnarray}
\{ Q_\alpha, Q_\beta \} \* 2\delta_{\alpha\beta} p_+ 
\equiv 2\delta_{\alpha\beta} H\ ,
\nonumber\\
\{ q_\sigma, Q_\beta \} \* 2\Gamma^i_{\sigma\beta} p_i\ ,
\nonumber\\
\{ q_\sigma, q_\rho \} \* 2\delta_{\sigma\rho} p_-  \ .
\end{eqnarray}
We can work in an eigenbasis for $p_-$ so that the $q_\sigma$ are
essentially oscillator variables.  The system can be separated
into center-of-mass variables $(q_\sigma,p_i)$ and the rest.
The algebra of the $q_\sigma$ determines how the center-of-mass
variables enter into $Q_\alpha$ and $H$ but does not otherwise
constrain the theory.  The essential problem is then
supersymmetric quantum mechanics with ${\cal
N}_Q=\frac{1}{2}{\cal N}$ supercharges $Q_\alpha$.  To count the
fermionic coordinates, consider first a single particle.  It
has $2^{{\cal N}/4}$ spin states, which can be thought of as
generated by the
$q_A$.\footnote{We are interested in
particles that are massless (or BPS) in $d$ dimensions, and so
transform in small multiplets.}
A state with
$k$ particles thus has $2^{k{\cal N}/4}$ states.  These would be
described by quantizing $N_F = \frac{1}{2} k{\cal N}$ real
fermionic coordinates, with action first order in time.
In table~1 we give the count of real bosonic and
fermionic coordinates in $3 \leq d \leq 10$, for the minimal
supersymmetry algebra in each dimension.
\begin{table}[t]
\caption{Number of real bosonic and fermionic coordinates per particle.}
\vspace{0.4cm}
\begin{center}
\begin{tabular}{|c|c|c|c|}
\hline
$d$&${\cal N}$ &$N_B/k$ &$N_F/k = {\cal N}_Q$ \\
 \hline
$3$&$2$ &$1$ &$1$ \\
$4$&$4$ &$2$ &$2$ \\
$5$&$8$ &$3$ &$4$ \\
$6$&$8$ &$4$ &$4$ \\
$7$&$16$ &$5$ &$8$ \\
$8$&$16$ &$6$ &$8$ \\
$9$&$16$ &$7$ &$8$ \\
$10$&$16$ &$8$ &$8$ \\
 \hline
\end{tabular}
\end{center}
\end{table}

Consider the case of $d=6$, where there are four
supercharges.  Supersymmetric QM with ${\cal N}_Q = 4$ can be obtained
by dimensional reduction from the familiar $D=4$, $N=1$ nonlinear
sigma models,\cite{zumino} giving QM on a K\"ahler
space.  A space with $4k$ bosonic coordinates can be described
with $2k$ chiral superfields, each having complex scalar. 
Each chiral superfield also contains two complex fermion
fields, giving a total of $8k$ real fermion fields.  Under
dimensional reduction to QM, each field becomes a coordinate.
From table~1, the number of fermionic coordinates in the
K\"ahler models is twice that in the DLCQ theory, so the latter are not
of the familiar type.  Note from the table that in dimensions 3, 4, 6,
and 10 the numbers of bosonic and fermion coordinates are equal.  In
other dimensions the number of fermionic coordinates is greater, but
always less than would be obtained with $D=4$, $N=1$ superfields.

After some work on this problem, we learned that the quantum
mechanical systems with $N_B = N_F$ (as in $d = 3, 4, 6, 10$) had already
been constructed in some detail by Gibbons, Papadopoulos, and
Stelle,\cite{GPS} following earlier work of 
Coles and
Papadopoulos,\cite{CP} Gibbons, Rietdijk, and van
Holten,\cite{GRV} and De Jonghe, Peeters, and Sfetsos ;\cite{DPS}
they have been further analyzed in a recent paper by Michelson and
Strominger.\cite{MS}  The physical motivation was different, namely
the study of black hole moduli spaces.  The QM systems for $d = 4,
6$, and $10$ are respectively the 2A, 4A, and 8A models of those
authors (for
$d=3$, where ${\cal N}_Q = 1$, the models are constructed with standard
superfields).  The supersymmetry multiplets cannot be obtained
by dimensional reduction from four dimensions, but they can be
obtained dimensional reduction of the
$(2,0)$, $(4,0)$, and $(8,0)$ multiplets in two dimensions.  Not
all quantum mechanical Lagrangians can be obtained by
dimensional reduction, because the one-dimensional theory has less
spacetime symmetry.

In this paper we will describe some small
extensions of the earlier work on these models.  First, we show how
the the 4A theories can be constructed with $D=1$, $N=4$ superfields
(the reduction of the familiar $D=4$, $N=1$ superfields) with an
extra constraint; the previous work used $D=1$, $N=1$ superfields. 
Second, we consider the addition of a gauge field on moduli space. 
Finally, we show how models with $N_B < N_F$ can be obtained by a
reduction of those with
$N_B = N_F$; a new feature here is that a potential energy term can
arise.

\section{Superfield Description of the 4A Theories}

To deduce the field content we consider the states of a free
massless particle in six dimensions.
The 6-dimensional supersymmetry algebra has an $SU(2)$
$R$-symmetry.  Under $SO(5,1)
\times SU(2)_R$ the supercharge transforms as $({\bf
4}, {\bf 2})$.  The DLCQ breaks this to
$SO(4) \times SU(2)_R = SU(2)_1 \times SU(2)_2 \times SU(2)_R$. 
The supercharge decomposes
\begin{eqnarray}
&(\frac{1}{2}, {\bf 2}, {\bf 1}, {\bf 2}):& Q_{AX} \nonumber\\
&(-\frac{1}{2}, {\bf 1}, {\bf 2}, {\bf 2}):& q_{MX} \ ,
\end{eqnarray}
where the $\pm \frac{1}{2}$ is the transformation under the
longitudinal $SO(1,1)$.  Indices $A, \ldots$, $M, \ldots$, and $X,
\ldots$ are used to label doublets of the three $SU(2)$'s.
The supercharges satisfy a reality condition
\begin{equation}
Q_{AX}^* = \epsilon^{AB} \epsilon^{XY} Q_{BY}\ , \qquad
q_{MX}^* = \epsilon^{MN} \epsilon^{XY} q_{NY}\ .
\label{real}
\end{equation}

On a massless particle with $p_+ = 0$, the $Q_{AX}$ vanish while
the $q_{MX}$ generate the spin states.  In the quantum
mechanics, the latter role is played by the quantized fermionic
coordinates $\psi$, so we can identify these as transforming as
$\psi_{MX}$ with the same reality condition as the $q_{MX}$.  The
bosonic coordinates are vectors of $SO(4)$ and so transform as
$X_{AM}$, again with a reality condition:
\begin{equation}
X_{AM}^* = \epsilon^{AB} \epsilon^{MN} X_{BN}\ , \qquad
\psi_{MX}^* = \epsilon^{MN} \epsilon^{XY} \psi_{NY}\ .
\label{realit}
\end{equation}
We can readily write down the
supersymmetry algebra for a free particle,
\begin{eqnarray}
[ Q_{AX}, X_{B}\!^{M} ] \* \sqrt{2} \epsilon_{AB} \psi^{M}\!_{X}\ ,
\nonumber\\ 
\{ Q_{AX}, \psi^{M}\!_{Y} \} \* i \sqrt{2} \epsilon_{XY} \dot
X_{A}\!^{M}\ .
\label{susy1}
\end{eqnarray}
We use conventions of Wess and Bagger.\cite{WB}  For example,
$\epsilon^{12} = \epsilon_{21} = +1$, and indices are raised and
lowered by left-multiplication with $\epsilon$.

Now let us compare with the superfields obtained by dimensional
reduction from $D=4$.  We should emphasize that dimensional
reduction and DLCQ, while they both lead to supersymmetric QM,
are completely different.  The DLCQ takes a quantum field theory
in $d$ dimensions to QM in $d-2$ spatial dimensions.  Reduction
takes quantum field theory in $D$ dimensions to quantum field
theory in 1 dimension (time), where the fields are reinterpreted
as coordinates.

To get the right number of bosons we take two chiral superfields
$\Phi^i$.  In addition to the doublet index $i$ there is a
doublet index $\alpha$ on the superspace coordinates $\theta$
and a doublet index $\dot\alpha$ on their conjugates.
Because the reduction breaks $SO(3,1)$ to $SO(3)$, indices
${}_\alpha$ and ${}^{\dot\alpha}$ transform in the same way.  The
superderivative algebra reduces to 
\begin{equation}
\{ D_\alpha, D_{\dot\beta} \} = 2i
\delta_{\alpha\dot\beta}\frac{\partial}{\partial t}
\ .
\end{equation}
The supersymmetry transformations of the scalar components are
\begin{eqnarray}
[ Q_{\alpha}, A^{i} ] \* \sqrt{2}
\psi^i_\alpha\ ,
\nonumber\\ 
{}[ Q_{\alpha}, A^{i*} ] \* 0\ ,
\end{eqnarray}
and the conjugate relations.
The DLCQ and superfield transformations are
of the same form provided we identify
\begin{equation}
Q_\alpha = Q_{2X} \big|_{X=\alpha} \ ,\quad  A^i =
X_{1}\!^{M}\big|_{M=i}\ ,
\quad \psi^i_\alpha = \psi^{M}\!_{X} \big|_{X=\alpha, M=i}
\ .  \label{dict}
\end{equation}
Notice in particular that $A^{i*}$ is then $-\epsilon_{ij}
X_{2}^j$.

From the introduction we know that the superfield description has twice
as many fermion fields as the DLCQ.  Here this arises because the
latter satisfy the reality condition,
\begin{equation}
\psi^{i*}_\alpha = -\epsilon_{ij} \epsilon^{\alpha\beta}
\psi^j_\beta\ .\label{psireal}
\end{equation}
This can be incorporated into the superfield formalism by the
new constraint
\begin{equation}
\bar D_{\dot\alpha} \Phi^{i*} = -\epsilon_{ij}
\delta_{\dot\alpha\alpha}
\epsilon^{\alpha\beta} D_{\beta} \Phi^j\ , \label{newcon}
\end{equation}
which is in addition to the usual chiral constraint $\bar D_{\dot\beta}
\Phi^j = 0$.  This has no effect on the lowest, bosonic, components
$A^i$, while eliminating half of the fermionic components.
It also eliminates the auxiliary components in terms of the velocities.

To describe $k$ free particles one would use $k$ pairs of superfields,
with the constraint~Eq.\ \ref{newcon} on each.  To obtain a
general interacting theory we must consider the generalization
\begin{equation}
\bar D_{\dot\alpha} \Phi^{i*} =
J^{\bar\imath}\!_{j\dot\alpha}\!^\beta (\Phi,\Phi^*)
D_{\beta} \Phi^j\ . \label{newgen}
\end{equation}
This nonlinear constraint must be consistent, meaning that it
represents only the same number of constraints as the linearized
version~Eq.\ \ref{newcon}, eliminating the auxiliary fields and
half the fermions, and leaving the lowest components
unconstrained. First, from the conjugate of Eq.~\ref{newgen} we
obtain
\begin{equation}
J^{\bar\imath}\!_{j\dot\alpha}\!^\beta (
J^{\bar\jmath}\!_{k\dot\beta}\!^\gamma)^* =
\delta^{\bar\imath}{}_{\bar k}
\delta_{\dot\alpha}{}^{\dot\gamma}
\end{equation}
Second, acting on both sides with $\bar D_{\dot\gamma}$ gives
\begin{equation}
\bar D_{\dot\gamma} \bar D_{\dot\alpha} \Phi^{i*} =
J^{\bar\imath}\!_{j\dot\alpha}\!^\beta{}_{,\bar k}
\bar D_{\dot\gamma} \bar\Phi^k D_{\beta} \Phi^j +
2i J^{\bar\imath}\!_{j\dot\alpha}\!^\beta
\delta_{\beta\dot\gamma} \partial_t
\Phi^j
\ . \label{barD}
\end{equation}
Focus on the lowest component of this equation.
The LHS is antisymmetric in $\dot\alpha\dot\gamma$; in order that
this equation not constrain the velocities, but just eliminate the
auxiliary field, the second term on the RHS must also be antisymmetric
in $\dot\alpha\dot\gamma$.  Thus, 
\begin{equation}
J^{\bar\imath}\!_{j\dot\alpha}\!^\beta(\Phi,\Phi^*)
= J^{\bar\imath}{}_{j}(\Phi,\Phi^*)
\delta_{\dot\alpha\alpha} \epsilon^{\alpha\beta}\ ,
\qquad J^{\bar\imath}{}_{j} J^j{}_{\bar k}
 = -\delta^{\bar\imath}{}_{\bar k} \label{almost}
\end{equation}
where $J^j{}_{\bar k} \equiv (J^{\bar \jmath}{}_{k})^*$.
The part of Eq.~\ref{barD} which is symmetric in
$\dot\alpha\dot\gamma$ comes only from the first term on the RHS,
and can now be written
\begin{equation}
0 = J^{\bar\imath}{}_{j, \bar k} 
J^{\bar k}{}_{l}(
D^{\gamma} \Phi^l D^{\alpha} \Phi^j + D^{\alpha} \Phi^l D^{\gamma}
\Phi^j) 
\label{barD1}
\end{equation}
The expression in parentheses is antisymmetric in $lj$. 
In order that Eq.~\ref{barD1} not represent new constraints on the
fields we need that it hold identically, and so
\begin{equation}
J^{\bar k}{}_{[l} J^{\bar\imath}{}_{j], \bar k} = 0\ . \label{barD2}
\end{equation}
Finally, acting on both sides of Eq.~\ref{newgen} with $D_\gamma$
and using the constraints gives
\begin{equation}
2i\epsilon_{\delta\gamma} \partial_t \Phi^{i*} =
J^{\bar\imath}{}_{j, k} D_{\gamma} \Phi^k
D_{\delta} \Phi^j
+ J^{\bar\imath}{}_{j}
D_{\gamma} D_{\delta} \Phi^j\ . 
\end{equation}
The antisymmetric part of this again relates the auxiliary field to the
velocities.  Only the first term on the RHS has a symmetric part, and
its vanishing identically implies
\begin{equation}
J^{\bar\imath}{}_{[j, k]}  = 0 \label{barD3}
\end{equation}

Eqs.\ \ref{almost},~\ref{barD2},
and~\ref{barD3} imply consistency of the constraints: for
example, one can regard them as determining the derivatives of
$\Phi_i$ with respect to
$\bar\theta^1$, $\bar\theta^2$, and $\theta^2$, and these are
integrable.  They imply
\begin{eqnarray}
\psi^{i*}_{1} \* J^{\bar\imath}{}_{j}(A, A^*) \psi^{j}_{2}
\nonumber\\ 
F^i \* -i J^i{}_{\bar \jmath} \dot A^{j*} + J^i{}_{\bar \jmath}
J^{\bar
\jmath}{}_{k,l}\psi^k_{1} \psi^l_{2}
\ , \label{constr}
\end{eqnarray}
for the components of \cite{WB} 
\begin{equation}
\Phi^i = A^i(y) + \sqrt{2} \theta\psi^i(y) + \theta\theta
F^i(y)\ ,
\end{equation}
where $y = t - i \theta^\alpha \bar\theta^{\dot\alpha}
\delta_{\alpha\dot\alpha}$.

The geometric interpretation of the
constraint~Eqs.\ \ref{almost},~\ref{barD2}, and~\ref{barD3} is as
follows.  Eq.\ \ref{almost} implies that the tensor
$I_2$ defined by
\begin{equation}
I_2 \biggl[ \begin{array}{c} \Phi^i \\ \Phi^{i*} \end{array}
\biggr] = \biggl[ \begin{array}{c} J^{i}{}_{\bar\jmath} \Phi^{j*}
\\  J^{\bar\imath}{}_{j} \Phi^j \end{array}
\biggr]
\end{equation}
is an almost complex structure.  This is in addition to the usual
complex structure
\begin{equation}
I_1 \biggl[ \begin{array}{c} \Phi^i \\ \Phi^{i*} \end{array}
\biggr] = \biggl[ \begin{array}{c} i\Phi^i \\ -i\bar\Phi^{i*}
\end{array}
\biggr]
\end{equation}
of the superfield formalism.  Eqs.~\ref{barD2} and~\ref{barD3} are
then the vanishing of the Nijenhuis tensor for $I_2$, written in the
complex coordinates associated to $I_1$.  Thus $I_1$ is a complex
structure, as is $I_3 = I_1 I_2 = - I_2 I_1$, in agreement with the
conditions found previously.\cite{GPS}

An invariant action is given by the superspace invariant
\begin{equation}
\int dt\,d^4\theta\,K(\Phi,\Phi^*)\ .
\end{equation}
This automatically satisfies the geometric conditions for
supersymmetry from refs.~\cite{GPS,MS} though we have not shown that all
solutions are of this form.\footnote{The superspace formalism is not
essential here; once the transformation laws are determined one can
take the fourth variation of a scalar function.}  The bosonic kinetic
term is
\begin{equation}
 K_{,i\bar\jmath}(A,A^*) (\dot A^i \dot A^{j*} + F^i F^{j*})
= K_{,i\bar\jmath}(A,A^*) (\dot A^i \dot A^{j*} + 
J^i{}_{\bar k} J^{\bar \jmath}{}_l \dot A^l \dot A^{k*} ) 
\label{kin}
\end{equation}
Introducing general real coordinates $x^a$, the metric on moduli space
can be written symmetrically among the three complex structures, 
\begin{equation}
g_{ab} = \sum_{r=0}^3 I_{r}{\!}^c{\!}_a I_{r}{\!}^d{\!}_b K_{,cd}\ ,
\label{genmet}
\end{equation}
where $I_0$ is the identity.

\section{Gauge Fields on Moduli Space}

A small generalization of previous work is to add a gauge field on
moduli space.  We will do this both with $D=1$, $N=1$ superfields, 
following ref.~\cite{GPS}, and the $D=1$, $N=4$ superfields above.

In $N=1$ superfields we add a term
\begin{equation}
S'=\int dt\,d\theta\, {\cal A}_a DX^a
\end{equation}
where the lowest component of $X^a$ is the real coordinate $x^a$.
Under the variation
\begin{equation}
\delta X^a = \epsilon I^a_{rb} DX^b\ , \label{n1susy}
\end{equation}
the variation of the action is
\begin{eqnarray}
\delta S'\*\int dt\,d\theta\, ( {\cal A}_{a,b} \delta X^b DX^a
+ {\cal A}_a D \delta X^a ) \nonumber
\\ \* \int dt\,d\theta\, ( {\cal A}_{a,b}
- {\cal A}_{b,a}) DX^b \delta X^a \nonumber
\\ \*  \epsilon\int dt\,d\theta\, {\cal F}_{ab} I^a_{rc} DX^b DX^c \ .
\end{eqnarray}
By antisymmetry of $DX^b DX^c$,
\begin{equation}
{\cal F}^{\vphantom a}_{ab} I^a_{rc} + {\cal F}^{\vphantom a}_{ca}
I^a_{rb} = 0\ .
\end{equation}
Thus the condition for 4A supersymmetry is that the indices of ${\cal
F}_{ab}$ be mixed --- ${\cal F}$ be a (1,1)-form --- in all three
complex structures.

In $N=4$ superfields consider the half-superspace integral
\begin{eqnarray}
S''\*\int dt\,d^2\theta\, W(\Phi,\Phi^*) + {\rm c.c.}
\nonumber\\
\* -i W_{,i} J^i{}_{\bar \jmath} \dot A^{j*} + {\rm fermionic} +
{\rm c.c.} \ .
\end{eqnarray}
so that
\begin{equation}
{\cal A}_{j} = i W_{,\bar\imath} J^{\bar\imath}{}_{j} \ ,\quad
{\cal A}_{\bar \jmath} = -i W_{,\bar\imath}^* J^i{}_{\bar \jmath} \
.\label{vect}
\end{equation}
We do not assume $W$ to be holomorphic, but will see how it is
constrained by supersymmetry.  The supersymmetry variation of $S''$ is
the integral of a total derivative except for the $\bar\theta$
derivative
\begin{equation}
\int dt\,d^2\theta\,  W_{,\bar\imath} D_{\dot\alpha} \Phi^{i*} + {\rm
c.c.} = -i \int dt\,d^2\theta\, 
{\cal A}_{j} D^{\alpha} \Phi^j + {\rm c.c.}
\end{equation}
using constraint Eq.~\ref{newgen}.  In order that this integrate
to zero, we must have ${\cal A}_{j} = \lambda_{,j}$ for some $\lambda$,
or equivalently
\begin{equation}
{\cal F}_{jk} = {\cal F}_{\bar\jmath\bar k} = 0 \label{fmixed1}
\ .
\end{equation}
Thus $\cal F$ is (1,1) with respect to $I_1$.  Eq.\ \ref{vect} is
equivalent to
\begin{equation}
0 = \partial_{[k} (J_{i]}{}^{\bar \jmath} {\cal A}_{\bar \jmath})
= J_{[i}{}^{\bar \jmath} \partial_{k]}{\cal A}_{\bar \jmath}\ ,
\label{amixed}
\end{equation}
using Eq.~\ref{barD3}.  Together with its conjugate this implies
\begin{equation}
J_{[i}{}^{\bar \jmath} {\cal F}_{k]\bar \jmath} = 0\ , \label{fmixed}
\end{equation}
which is the statement that $\cal F$ is (1,1) with respect to both
$I_2$ and $I_3$.  Thus we recover the same models as with $N=1$
superfields.\footnote{Eq.\ \ref{amixed} would
appear to be stronger than~Eq.\ \ref{fmixed}, but the
difference is a gauge choice. The earlier~Eq.\ \ref{fmixed1}
implies that
${\cal A}_j = \partial_j \lambda$ and ${\cal A}_{\bar\jmath} =
\partial_{\bar\jmath}(\lambda^*)$.  The real part of $\lambda$ can be
set to zero by a gauge transformation, and then~Eqs.\
\ref{amixed} and~\ref{fmixed} are equivalent.}

\section{Dimensional Reduction}

Systems with $N_B < N_F$ can be obtained from
those with $N_B = N_F$ by a process of reduction.  This requires
that the original system have one or more isometries.  With a
single isometry, for example, we can choose coordinates in which it
takes the form $\delta \xi =
\epsilon$.    Then $\xi$ does not appear in the action
undifferentiated, and so $\dot
\xi$ can be regarded as an independent auxiliary field $F$.

In $N=1$ superfield language, the corresponding superfield $\Xi$
does not appear in the action but $D \Xi$ may.  The lower
component of
$D \Xi$ is a fermion and the upper is $\dot \xi \to F$.  Such
superfields, with a fermion and an auxiliary field, were
considered in ref.~\cite{CP}, where they were denoted $\psi$. 
They can be used to construct models with $N_F > N_B$. However,
it appears that not all such models can be obtained
by dimensional reduction: we can work backwards, introducing a
superfield $\Xi$ and replacing $\psi$ everywhere with
$D\Xi$, only if the supersymmetry variations of $\psi$ are total
spinor derivatives.  For $D=1$, $N=1$ systems this is always
possible, but for higher
supersymmetries it is nontrivial.

This reduction process is suggested by the application to 
DLCQ: the systems in table~1 having $N_B < N_F$ can be obtained
from those with $N_B = N_F$ by dimensional reduction.  For example
we can go from the $d=6$ system to the $d=5$ system by
Kaluza-Klein reduction on $x^4$.  In a $k$-particle system, all
wavefunctions will be independent of $x^4$, so there are actually
$k$ isometries.

\subsection{Examples}

Let us illustrate the reduction for the 4A theory with a
single isometry.
The general bosonic action with an isometry is
\begin{equation}
\frac{1}{2} g_{mn} \dot x^m  \dot x^n + \frac{1}{2}
g_{\xi\xi} (\dot \xi + {\cal A}'_m \dot x^m)^2 + {\cal A}_m \dot
x^m + {\cal A}_\xi \dot \xi\ .
\end{equation}
Here $m,n$ run only over the $4k-1$ non-cyclic coordinates.  Note
the Kaluza-Klein gauge field on moduli space ${\cal A}'_m$, which
is in addition to the gauge field $({\cal A}_m, {\cal A}_4)$
introduced in section~4.  Replacing
$\dot
\xi \to F$ and solving for $F$ gives
\begin{equation}
\frac{1}{2} g_{mn} \dot x^m  \dot x^n - \frac{1}{2
g_{\xi\xi} } {\cal A}^2_\xi + ({\cal A}_m 
- {\cal A}_\xi {\cal A}'_m) \dot x^m \ .
\end{equation}
Thus a potential energy ${\cal A}^2_\xi / 2
g_{\xi\xi}$ is introduced.  The net reduced
gauge field
${\cal A}_m  - {\cal A}_\xi {\cal A}'_m$ is invariant under
reparameterization of $\xi \to \xi + \lambda(x^m)$.

Now let us begin with the simplest example, 
$k=1$ with the linear constraint~Eq.\ \ref{newcon} and no gauge
field.   The bosonic action in~Eq.\ \ref{kin} takes the form
\begin{equation}
(K_{,1\bar 1} + K_{,2\bar 2}) (\dot A^1 \dot A^{1*} + \dot A^2
\dot A^{2*})\ .
\end{equation}
In real coordinates this is
\begin{equation}
\frac{1}{2} H(x^a)\, \dot x^b \dot x^b\ ,
\end{equation}
where $H = \nabla^2 K$ is a generic function of $x^1,x^2,x^3,x^4$. 
This example appears in refs.~\cite{chs,GPS}.

Now reduce under translation in $x^4$.  To make the result more
interesting add a constant potential ${\cal A}_4 = C_1$.
Then $H$ depends only on $x^m \equiv x^1,x^2,x^3$ and the reduced
bosonic action is
\begin{equation}
\frac{1}{2} H(x^m)\, \dot x^n \dot x^n -
\frac{C_1^2}{2H(x^m)}  \ .
\end{equation}

Focusing now on systems with spherical symmetry we have
$H = H(r)$.  An obvious generalization is to add a spherically
symmetric magnetic field
\begin{equation}
{\cal B}_m = \frac{C_2 x^m}{r^3}\ ,\qquad {\cal A}_4 = C_1 +
\frac{C_2}{r} \ .
\end{equation}
The form of ${\cal A}_4$ follows from the (1,1) condition on
${\cal F}$, which here reduces to self-duality.  The reduced
action is then
\begin{equation}
\frac{1}{2} H(r)\, \dot x^n \dot x^n -
\frac{1}{2H(r)} \biggl( C_1 + \frac{C_2}{r} \biggr)^2
+ {\cal A}_m \dot x^m \ . \label{spher2}
\end{equation}

An independent search for spherically symmetric quantum
mechanical systems produced a further one-parameter generalization
of these models.  The bosonic actions are all of the form
Eq.~\ref{spher2}; the generalization appears in the supersymmetry
transformations and consequently the fermionic terms.  These
models can be obtained from reduction as follows.  For $k=1$ there
is a one-parameter generalization of the complex structures, given
by Taub-NUT space, with the constant complex structure as a limit. 
This space has an $SU(2) \times U(1)$ isometry, so reducing on the
$U(1)$ leaves a theory with $SU(2)$ rotational symmetry.

\subsection{Taub-NUT Reduction}

In the remainder of this paper we work out some of the
(rather tedious) details of this reduction, beginning with the
bosonic action.  In coordinates $(x^1,x^2,x^3,\xi)$ with $\xi$
cyclic, the three complex structures are~\cite{taubnut}
\begin{eqnarray}
I_{1}\!^a\!_b \* \frac{1}{S} \left[ \begin{array}{cccc}
-a_1 & -a_2 & 0 & -1 \\
0 & 0 & -S & 0 \\
0 & S & 0 & 0 \\
S^2 + a_1^2 & a_1 a_2 & S a_2 & a_1 \end{array} \right]
\nonumber\\
I_{2}\!^a\!_b  \* \frac{1}{S} \left[ \begin{array}{cccc}
0 & 0 & S & 0 \\
-a_1 & -a_2 & 0 & -1 \\
-S & 0 & 0 & 0 \\
a_1 a_2 &  S^2 + a_2^2 & - S a_1 & a_2  \end{array} \right]
\nonumber\\
I_{3}\!^a\!_b  \* \left[ \begin{array}{cccc}
0 & -1 & 0 & 0 \\
1 & 0 & 0 & 0 \\
0 & -S^{-1} { a}_2 & 0 & - S^{-1} \\
- { a}_2 & { a}_1 & S & 0 \end{array} \right]\ ,
\end{eqnarray}
where $m$ is the Taub-NUT parameter and $a_n$ is a magnetic monopole
field of unit strength,
\begin{equation}
S = \frac{1}{2m} + \frac{1}{r}\ ,\qquad a_\phi = \cos \theta\ .
\end{equation}
After some computation, one finds from
Eq.~\ref{genmet} that the general metric with a $U(1)$ isometry is
\begin{equation}
ds^2 = H(x^m) \bigl[ d x^n d x^n + S^{-2}(d \xi + a_n
d x^n )^2 \bigr] \ , 
\end{equation}
where $H = \partial_m \partial_m K$ and $K$ is independent of
$\xi$.  For $H = 2mS$ this is the hyper-K\"ahler Taub-NUT metric,
so these metrics are conformal to Taub-NUT.

We now add a gauge field ${\cal A}_a$.  We find it convenient to
impose rotational invariance before the $(1,1)$
condition.  The condition that the gauge field term ${\cal A}_a
\dot x^a$ be invariant up to a time derivative under a general
Killing vector $L^a$ is
\begin{equation}
L^a {\cal F}_{bc,a} = L^a\!_{,b} {\cal F}_{ca} 
- L^a\!_{,c} {\cal F}_{ba} \ .
\end{equation}
In this case the Killing vectors are
\begin{eqnarray}
L_1 \* x^2 \partial_3 - x^3 \partial_2 + \frac{r x^1}{(x^1)^2 +
(x^2)^2}\partial_\xi \ ,\nonumber\\
L_2 \* x^3 \partial_1 - x^1 \partial_3 + \frac{r x^2}{(x^1)^2 +
(x^2)^2}\partial_\xi\ ,\nonumber\\
L_3 \* x^1 \partial_2 - x^2 \partial_1 \ ,\quad L_\xi =
\partial_\xi\ .
\end{eqnarray}
Rotational invariance then implies
\begin{eqnarray}
{\cal F}_{m\xi} = x^m f(r)\, ,\quad
{\cal F}_{12} = x^3 g(r)\, ,\quad
{\cal F}_{23} = x^1 h(r,x^3)\, ,\quad {\cal F}_{31} = x^2
h(r,x^3)\ ,
\end{eqnarray}
where $h(r,x^3) = g(r) - r f(r) / (r^2 - (x^3)^2)$.  The Bianchi
identity becomes
\begin{equation}
f + rf' = 3rg + r^2 g'
\end{equation}
and the $(1,1)$ conditions reduce to
\begin{equation}
g = -f/2m\ .
\end{equation}
Integrating the Bianchi identity gives 
\begin{equation}
f = \frac{a}{r(r+2m)^2}
\end{equation}
with $a$ an integration constant.  The gauge field is then
\begin{equation}
{\cal A}_\phi = \frac{az}{2m(r+2m)}
\ ,\qquad {\cal A}_\xi = - \frac{a}{r+2m} + b\ .
\end{equation}

Upon reduction, the bosonic action is the same as in
Eq.~\ref{spher2} with
\begin{equation}
C_1 = \frac{b}{2m}\ ,\qquad C_2 = \frac{2mb-a}{2m}\ .
\end{equation}
Taking the limit $m \to 0$ with $C_1$ and $C_2$ fixed gives the
earlier models.
In the bosonic action the three parameters $(m, a, b)$ collapse into
two,
$(C_1,C_2)$.  However, the third parameter appears in the
supersymmetry transformation, Eq.~\ref{n1susy}, since this depends
on the complex structure and therefore on $m$ separately.  The
fermionic terms in the action are therefore also affected.  The
general Lagrangian can be written
\begin{eqnarray}
&&\frac{1}{2} g_{ab} \dot x^a \dot x^b + \frac{i}{2} \psi^a
\bigl[ g_{ab} \dot \psi^b + (\Gamma_{abc} - c_{abc}) \dot x^b
\psi^c \bigr] - \frac{1}{6} c_{abc,d} \psi^a  \psi^b \psi^c
\psi^d \nonumber\\
&& \qquad\qquad\qquad\qquad\qquad\qquad
{}+ {\cal A}_a \dot x^a + \frac{1}{2} {\cal F}_{ab}
\psi^a \psi^b\ ,
\end{eqnarray}
where $c$ is a torsion form.\cite{GPS,MS}  The torsion form
here is 
\begin{equation}
c = \biggl(H'(r) + \frac{2m}{r^2} \biggr) \sin \theta d\phi \wedge
d\theta \wedge d\xi
\end{equation}
(see the appendix to ref.~\cite{chs} or Eq.~4.14 of
ref.~\cite{GPS}).  Thus the fermionic terms depend separately on
$m$, through $g_{ab}$ and $c_{abc}$.  For the Taub-NUT metric the
torsion vanishes, as it must.

\section*{Acknowledgments}
This work
was supported in part by NSF grants PHY94-07194 and PHY97-22022.

\end{document}